\documentclass[sigconf,nonacm]{aamas}

\usepackage{balance} 

\usepackage{graphicx}
\usepackage{cleveref}
\usepackage{wrapfig}
\usepackage{amsmath}
\usepackage{caption}
\usepackage{subcaption}
\usepackage{dsfont}
\usepackage{amsmath}
\DeclareMathOperator*{\argmax}{arg\,max}

\newcommand{\leader}{L}
\newcommand{\followers}{\mathbf{F}}

\setcopyright{ifaamas}
\acmConference[AAMAS '25]{Proc.\@ of the 24th International Conference
on Autonomous Agents and Multiagent Systems (AAMAS 2025)}{May 19 -- 23, 2025}
{Detroit, Michigan, USA}{A.~El~Fallah~Seghrouchni, Y.~Vorobeychik, S.~Das, A.~Nowe (eds.)}
\copyrightyear{2025}
\acmYear{2025}
\acmDOI{}
\acmPrice{}
\acmISBN{}

\acmSubmissionID{102}

\title[ADAGE]{ADAGE: A generic two-layer framework for adaptive agent based modelling}

\author{Benjamin Patrick Evans}
\affiliation{
  \institution{JP Morgan AI Research}
  \city{London}
  \country{United Kingdom}}
\email{benjamin.x.evans@jpmorgan.com}

\author{Sihan Zeng}
\affiliation{
  \institution{JP Morgan AI Research}
  \city{Palo Alto}
  \country{USA}}
\email{}

\author{Sumitra Ganesh}
\affiliation{
  \institution{JP Morgan AI Research}
  \city{New York}
  \country{USA}}
\email{}

\author{Leo Ardon}
\affiliation{
  \institution{JP Morgan AI Research}
  \city{London}
  \country{United Kingdom}}
\email{}

\begin{abstract}

Agent-based models (ABMs) are valuable for modelling complex, potentially out-of-equilibria scenarios. However, ABMs have long suffered from the Lucas critique, stating that agent behaviour should adapt to environmental changes. Furthermore, the environment itself often adapts to these behavioural changes, creating a complex bi-level adaptation problem. Recent progress integrating multi-agent reinforcement learning into ABMs introduces adaptive agent behaviour, beginning to address the first part of this critique, however, the approaches are still relatively ad hoc, lacking a general formulation, and furthermore, do not tackle the second aspect of simultaneously adapting environmental level characteristics in addition to the agent behaviours.
In this work, we develop a generic two-layer framework for ADaptive AGEnt based modelling (ADAGE) for addressing these problems. This framework formalises the bi-level problem as a Stackelberg game with conditional behavioural policies, providing a consolidated framework for adaptive agent-based modelling based on solving a coupled set of non-linear equations.
We demonstrate how this generic approach encapsulates several common (previously viewed as distinct) ABM tasks, such as policy design, calibration, scenario generation, and robust behavioural learning under one unified framework. We provide example simulations on multiple complex economic and financial environments, showing the strength of the novel framework under these canonical settings, addressing long-standing critiques of traditional ABMs.

\end{abstract}

\newcommand{\BibTeX}{\rm B\kern-.05em{\sc i\kern-.025em b}\kern-.08em\TeX}

\begin{document}


\pagestyle{fancy}
\fancyhead{}


\thanks{Accepted at the
2025 International Conference on Autonomous Agents and Multiagent Systems (AAMAS)}

\maketitle 


\section{Introduction}

Agent-based models (ABMs) have shown great promise for modelling scenarios where outcomes may differ from traditional equilibrium analysis \cite{axtell2022agent}, for example, providing more realistic models of the dynamics in complex systems where deriving strict equilibria is unrealistic or impractical \cite{arthur2021foundations}.

However, traditional ABMs crucially rely on the behavioural rules of the agents, which are typically fixed and/or manually specified. This fixed behaviour opens ABMs to the famed \textit{Lucas Critique}:
\begin{quotation}
\textit{Given that the structure of  a model consists of optimal decision rules of agents, and that optimal decision rules vary systematically with changes in the environment, it follows that any change in policy will systematically alter the structure of the models} -- paraphrased from \cite{lucas1976econometric}
\end{quotation}
\noindent 
raising concerns about using fixed behaviour rules in ABMs \cite{napoletano2018short}. 

Such a critique can be addressed with the introduction of adaptive agent behaviour \cite{turrell2016agent}, e.g. by using Artificial Intelligence (AI) techniques to introduce models of heterogeneous agents adapting their behaviour \cite{ardon2023phantom,aymanns2018models}
following the literature on adaptive learning in macroeconomics \cite{hommes2022canvas}. For example, 
\cite{salle2015modeling} models adaptive expectations in response to economic environment changes, \cite{catullo2022forecasting} lets firms forecast sales based on news (e.g. shocks), and \cite{zheng2022ai} model household behaviour in response to changing macroeconomic policies. However, this is just one part of the adaptive problem. Crucially, the environment itself also often adapts to these behavioural changes (e.g., macroeconomic policy changing based on the agent behaviour), which has received comparatively little attention \cite{zheng2022ai}.

In this work, we develop a generic two-layer framework for ADaptive AGEnt-based modelling (ADAGE). ADAGE automatically learns behavioural rules that adapt to environmental changes, as well as learning these potential environmental changes based on the task of the designer (e.g., for policy design or calibration). The framework is built around a bi-level design, with an inner simulation layer, with agents learning behavioural rules conditioned on an observation of updateable characteristics $\boldsymbol{\theta}$, and an outer layer updating these characteristics $\boldsymbol{\theta}$. As agent behaviour is conditioned on the evolving characteristics, this helps mitigate the Lucas critique, adapting behaviour in response to environmental change, and, due to the bi-level design, simultaneously optimises the environment itself in response to the changing agent behaviours. We formulate the problem as a Stackelberg game, solving a coupled set of non-linear equations, and show how this subsumes several common ABM tasks, such as calibration, scenario analysis, robust behavioural learning, and policy design, under one unified framework.

Specifically, we:
\begin{itemize}
    \item Develop a generic framework for adaptive agent-based modelling based on Stackelberg games
    \item Derive several common ABM tasks (such as calibration and policy design) as special instances of the framework
    \item Evaluate the framework showing the flexibility and suitability in complex agent-based models.
\end{itemize}

The proposed approach is a generic framework for developing adaptive simulations --- learning and adjusting agent behaviour \textit{and} environment parameters in a tractable manner.

\section{Background and Related Work}

The combination of machine learning techniques (such as Reinforcement learning) with agent-based simulation is a growing research area with much promise for improving upon hand-crafted models \cite{tilbury2022reinforcement,an2021challenges}, while maintaining some of the key benefits such as agent heterogeneity and relaxed notions of equilibria \cite{evans2024learning}.

However, most existing work only considers the learning of fixed behavioural rules for agent-based simulation \cite{9552052,vargas2023deep,sert2020segregation,dehkordi2023using,klugl2023modelling}, requiring retraining under novel conditions. For example, in a market setting, if the macroeconomic policies such as taxation rates changed, this would require retraining the model rather than automatically modelling the adaption in agent behaviour based on the updated tax rates. This issue was highlighted by Geanakoplos \cite{geanakoplos2012getting}, where following regulatory policy changes, required redefining agent behaviours in their pioneering housing ABM \cite{geanakoplos2012getting}. A more advanced two-layer approach\footnote{Despite similarities in terminology, the approach is distinct from Multi-level agent-based models (MLM) \cite{hjorth2020levelspace}. MLM refers to the various levels of granularity/abstractions present in simulation (e.g. individuals, suburbs, cities, countries). In contrast, we use the term two-layer framework to refer to an outer and inner simulation layer.} for not only learning agent behaviour but adapting this behaviour in response to changing macroeconomic conditions is introduced in the AI Economist \cite{zheng2022ai,trott2021building}, providing an approach for AI-driven tax policy design where agent behaviours adapt automatically. The AI Economist features an inner layer (for simulation) and an outer layer (for policy design), serving as the key motivation for this work. While the results of \cite{trott2021building} are extremely impressive, the framework is posed for the specific problem of AI-driven tax policy design and lacks a more general formulation for other common modelling tasks.

In this work, we develop a generic expansion based on the ideas of \cite{zheng2022ai} for adaptive agent-based modelling, unifying several previously distinct ABM tasks, something yet to be considered. For example, in addition to the policy design task of \cite{zheng2022ai}, we show how other common tasks, such as model calibration and scenario generation, can be encapsulated under a more general framework, with the AI economist then serving as a specific instantiation of the generic framework developed. This generic expansion is based on formulating these two-layer frameworks as Stackelberg games, which are asymmetric Markov games between a leader and one or more followers \cite{brero2022stackelberg, gerstgrasser2023oracles}. This formulation has been successfully investigated in the context of economic games \cite{brero2022learning}, but has not yet been thoroughly investigated for complex agent-based models. Therefore, in the following section, we formulate adaptive agent-based model design as a Stackelberg game, with the leader as the outer layer, and the followers as the (inner) agent-based simulation layer. 


\section{Proposed Approach}

\begin{figure}[!htb]
  \centering
  \includegraphics[width=.95\columnwidth]{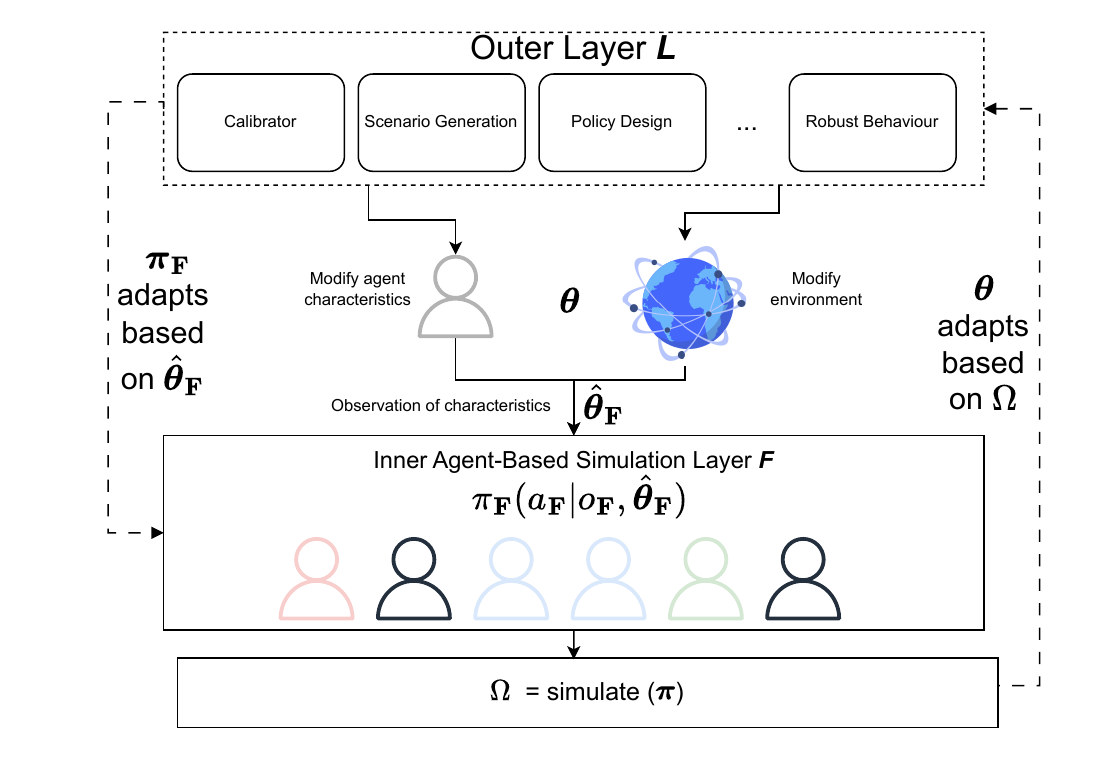}
  \caption{ADAGE: Two-layer framework.}\label{figFramework}
\end{figure}

We propose a generic two-layer framework, ADAGE, as visualised in \cref{figFramework}. This framework is represented as a Stackelberg game between a leader in the outer layer and $n$ followers in the inner simulation layer,
operating in  a parameterised environment (see parametric games \cite{fiscko2019control,zhang2024computing}) representing the agent-based model. 

\subsection{Formulation}

We model the problem as a Partially Observable Markov Game (POMG) with $n+1$ agents \cite{gerstgrasser2023oracles} -- 
agent $0$ is reserved as the leader while the remaining $n$ agents are the followers\footnote{Alternative representations considering multiple leaders and relaxing equilibria could also be considered.}. 
$\leader=0$ indexes the leader agent, and $\followers=\{1,\dots,n\}$ denotes the set of follower agents, and we refer to them as the outer layer and inner simulation layer, respectively.
The game can be characterised by the tuple:
\begin{equation}
    (S, A, T, r, O, \gamma)
\end{equation}
where $S$ is the state space, $A = (A_0, A_1, \dots, A_n)$ the action space of the agents, $T: S \times A \rightarrow S$ the transition function, $r: S \times A \rightarrow \mathbb{R}^N $ the reward functions, and $O = (O_0, O_1, \dots, O_N)$ the observation spaces, and $\gamma$ the discount rate.
The state $s\in S$ of the Markov game may not be generally known to any agent. Instead, agent $i$ has a local observation of the state $o_i(s)$ where $o_i:S\rightarrow O_i$.


Stackelberg games are essentially Markov games with inherent asymmetry, in which 1) the outer layer $\leader$ usually exerts a stronger effect on the state transition/reward and has more informative observations, 2) the outer layer $\leader$ and inner layer $\followers$ may operate on separate decision-making timescales $t_\leader$, $t_\followers$ (where typically $t_\leader > t_\followers$), and 3) the objectives of $\leader$ and $\followers$ exhibit hierarchical structure.
The outer layer acts first, with the goal of maximising a global objective, and the inner layer reacts, with each follower aiming to maximise its local objective given the leader's behaviour.

We characterise the effect of the leader's behaviour on the follower's behaviour through a characteristics variable $\boldsymbol{\theta}$. The characteristics variable is an outcome of $\pi_L$ and parameterises the environment perceived by the followers. Each follower $i\in\followers$ has a local observation $\hat{\boldsymbol{\theta}}_i$ of the characteristics information, which drives their behaviour.
We will discuss the meaning of $\boldsymbol{\theta}$ and $\hat{\boldsymbol{\theta}}_i$ in each specific task that we model. 

In the rest of the paper, we denote the behavioural policy of agent $i$ by $\pi_i(a\mid o_i,\hat{\boldsymbol{\theta}}_i)$ to emphasise that the followers condition their behavioural policies on the (observed) environment characteristics. When the conditioning is not explicit, it is assumed that $\hat{\boldsymbol{\theta}}_i$ is part of the observation $o_i$.

At timestep $t$ each active agent $i$ takes an action $a_{i,t} \sim \pi_i(o_{i,t})$ based on their behavioural policy $\pi_i$ and private observation $o_{i,t}$, receiving reward $r_{i,t}$ from the environment $\Omega$. The goal of agent $i$ is to find a behavioural policy $\pi_i$ to maximise their expected discounted return: 
\begin{equation}\label{eqExpectedDiscountedReturn}
R_i = \mathbb{E}\biggl[\sum_t \gamma^t r_{i,t} (s_t, a_{i, t}, a_{-i, t})\biggr]
\end{equation}
where $a_{-i, t}$ is the action of the other agents. The per timestep reward $r_{i,t}$ is generally based on the outputs of the agent-based simulation $\Omega$.

We denote the leaders behavioural policy as $\pi_\leader$, and the set of follower behavioural policies as $\pi_\followers=\{\pi_i:i\in\followers\}$. Similar notation is used
for the action $a$, reward function $r$, observation $o$, etc.
A Stackelberg equilibrium is a point where the leader maximises its expected return conditioned on followers reacting optimally:
\[\pi_\leader^* \in \argmax_{\pi_\leader} [R_\leader | \pi_\followers \sim \varepsilon(\pi_\leader)]\]
where $\varepsilon$ represents a best-response oracle (e.g. RL, Bayesian, or other optimisation function) \cite{gerstgrasser2023oracles}. Generally, the task is to simultaneously optimise $\pi_\leader$ and $\pi_\followers$ to solve this Stackelberg equilibria, as in practice the oracle $\varepsilon$ cannot be directly queried. The focus of this work is not on theoretical convergence guarantees to the equilibrium (which is analysed elsewhere \cite{pmlr-v119-fiez20a,zheng2022stackelberg}), instead we \textit{approximate} the equilibria by learning (approximate) best responses, and focus on its applications to adaptive agent-based modelling, irrespective of the particular optimisation algorithm used.

\subsubsection{Example tasks}

The framework's generality is in the outer layer's ability to serve multiple purposes based on the outcome of interest. By varying the leader's reward function and action space, we can capture many common ABM tasks. For example, if the desired outcome is policy or mechanism design \cite{zhang2024social}, the outer layer creates new rules, e.g. taxation rules \cite{zheng2022ai} through updating $\boldsymbol{\theta}$ (e.g., Social Environment Design \cite{zhang2024social}). If the goal is calibrating a simulator, the outer layer fits latent parameters $\boldsymbol{\theta}$ of agent characteristics to ensure the simulated outcomes most closely match the real-world dynamics \cite{vadori2022towards}. If the desired outcome is adaptability to unseen scenarios, the outer layer encourages robust behavioural policies in the spirit of meta-learning \cite{collins2022maml}. 

The rewards for the outer layer are based on outputs from the agent-based simulator $\Omega$. For example, for calibration, we can represent the leaders reward function as:
\begin{equation}\label{eqCalibration}
r_{\leader,t} = -|m(\Omega_t) - \varphi_t|
\end{equation}
where $m$ is some metric function from the environment $\Omega$ (e.g. resulting market prices), and $\varphi$ is the target metric (e.g. pricing from some ground truth data). The action  $a_{\leader,t}$ is then to tune calibratable latent parameters $\boldsymbol{\theta}$ to optimise \cref{eqCalibration}. Alternatively, we could think of the case of policy design where the goal of the leader is to improve social welfare for the follower agents:
\[
r_{\leader,t} = \mathbb{E}\sum_{i\in\followers}r_{i,t}
\]
or other more advanced social welfare functions \cite{zhang2024social}. Likewise, an adversarial leader could be represented with:
\[
r_{\leader,t} = -\mathbb{E}\sum_{i\in\followers}r_{i,t}
\]
in each case updating the parameters $\boldsymbol{\theta}$ of the ABM in an effort to improve $R_\leader$. Follower behaviour is then conditioned on observations of these characteristics $\pi_\followers(a | o_\followers, \hat{\boldsymbol{\theta}}_\followers)$, guiding the system behaviour based on optimising the rewards of interest.

The flexibility of ADAGE is in defining different reward structures and action spaces for the leader agent to achieve various tasks while maintaining the same representation (POMG) and solution concept (Stackelberg equilibria) for the leader-follower dynamics. In \cref{secExperiments} we provide concrete examples of this generality by subsuming several tasks under the framework.

\subsection{Optimisation}



The objective in optimising $\pi_\leader$ and $\pi_\followers$ is to find a Stackelberg equilibrium, i.e. a solution $(\pi_\leader^*,\pi_\followers^*)$ at which no agent can improve its local objective function holding the behaviour of all other agents fixed. Due to the ``gradient domination'' condition satisfied by \cref{eqExpectedDiscountedReturn}, every stationary point is globally optimal \cite{agarwal2021theory}. This implies that to find $(\pi_\leader^*,\pi_\followers^*)$, it suffices to solve the following coupled system of $n+1$ non-linear equations. Each equation $i$ states that the behavioural policy $\pi_i$ is a first-order stationary point for agent $i$ given that every other agent $j\neq i$ follows $\pi_j$
\begin{equation}
\left\{\quad
\begin{aligned}
&\nabla_{\pi_i^*}R_i=0,\quad \forall i\in \followers,\\
&\nabla_{\pi_\leader^*}R_{\leader}=0.
\end{aligned}
\right.\label{eq:system_equations}
\end{equation}
Note that in cases of large $n$, in order to reduce the number of equations that must be solved, shared policy learning \cite{vadori2020calibration} can be used for agents of similar types, reducing the set of follower behavioural policies to a subset ($<<n$) of shared behavioural policies.

The standard method, and the one we generally use here, for solving such coupled equations is alternating gradient descent/ascent (A-GD), i.e. we maintain parameter $\pi_{i,t}$ as an estimate of $\pi_i^*$ and take turns to update $\pi_{i,t}$ for every agent $i$ in the direction of $\tilde{\nabla}_{\pi_{i,t}}R_i$, where $\tilde{\nabla}_{\pi_{i,t}}R_i$ denotes an estimate of the exact gradient $\nabla_{\pi_{i,t}}R_i$ with the behaviour of agents $j\neq i$ fixed to their latest iterates:
\begin{equation*}
\begin{aligned}
&\pi_{\leader,t+1}=\pi_{\leader,t}+\alpha_{\leader,t} \tilde{\nabla}_{\pi_{i,t}}R_\leader\text{ given $\{\pi_{j,t}\}_{j\in \followers}$}\\
&\pi_{i,t+1}=\pi_{i,t}+\alpha_{i,t} \tilde{\nabla}_{\pi_{i,t}}R_i\text{ given $\{\pi_{j,t}\}_{j\neq i}$},\;\forall i\in \followers
\end{aligned}
\end{equation*}

When $R_i$ exhibits strong structure,
the convergence of the A-GD algorithm to $(\pi_\leader^*,\pi_\followers^*)$ is guaranteed under proper learning rates $\alpha_{\leader,t}$, $\alpha_{\followers,t}$ \cite{hong2023two,zeng2024two}. In general, we need to choose $\alpha_{\leader,t}$ to be much larger than all $\alpha_{\followers,t}$ to approximate a nested-loop algorithm that runs multiple inner layer updates per outer layer update.

\subsubsection{Conditional behavioural policies}

A key component of the proposed approach is learning follower behavioural policies $\pi_\followers$ conditioned on observations $\hat{\boldsymbol{\theta}}_\followers$ of the environment characteristics $\boldsymbol{\theta}$. While in practice many learning algorithms could be used for finding $\pi_\followers(a | o_\followers, \hat{\boldsymbol{\theta}}_\followers)$, and the framework is independent of any particular algorithm, here we use deep reinforcement learning (DRL) (specifically, PPO \cite{schulman2017proximal}), treating  $\hat{\boldsymbol{\theta}}_\followers$ as part of the observation space. The benefit of using DRL is that it natively parameterises the decision function by the agents observation $o_i$, permitting any functional form for the behavioural rule and removing the requirement of manually determining the rule or specifying the form of the decision function \textit{a priori}. 

\section{Experiments}\label{secExperiments}

In order to demonstrate the flexibility of the framework, we provide four illustrative (but non-exhaustive) examples of different outer layer configurations spanning distinct modelling tasks, showing how each task can be viewed as a special instance of the proposed framework. For each configuration, we additionally use a separate agent-based model in the inner simulation layer to demonstrate that the framework is simulator-independent, as well as using a combination of individual and shared policy learning. We generally use reinforcement learning to learn the behaviour in both layers, but we show that the framework is learning algorithm-independent, as demonstrated with alternative configurations such as Bayesian layers and analytically derived behavioural rules. The chosen environments are well established, spanning several economic and financial domains. Additionally, the environments feature a mixture of continuous and discrete action and observation spaces, showing the generality of the approach with many different ABMs.  The source code is provided in the supplementary material.

\subsection{Policy Designer}\label{secPolicyDesign}

\begin{figure}[!htb]
    \centering
    \includegraphics[width=.5\columnwidth]{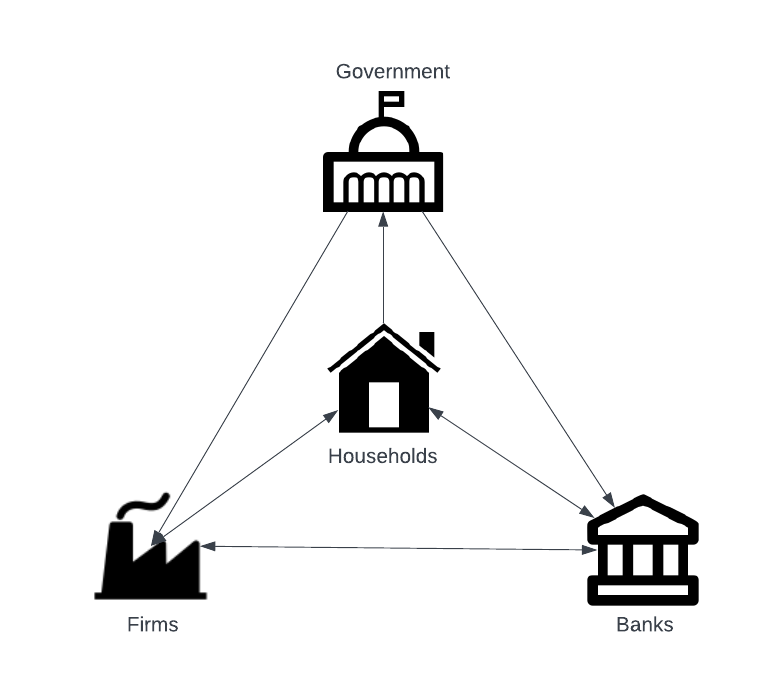}
    \caption{TaxAI: Economic Simulator}
    \label{figTaxAI}
\end{figure}

Policy design is an important use case for modelling, for example, helping to design COVID containment strategies in epidemiology \cite{quera2023don},
optimal auction design \cite{sandholm2003automated}, and assisting government and economic policy-making \cite{zheng2022ai,zhang2024social}. As we demonstrate below, policy design naturally fits within the ADAGE framework.

\subsubsection{Environment}

To demonstrate the use of ADAGE for policy design, we base an environment on TaxAI, "the most realistic economic simulator for optimal tax policy" \cite{mi2023taxai}, based on the canonical Bewley-Aiyagari model. Under this environment, there are multiple households, firms, banks, and a central government (visualised in \cref{figTaxAI}). This environment is similar to the one used in the AI-Economist \cite{zheng2022ai}, and ABIDES-Economist \cite{dwarakanath2024abides}. The government acts as the leader $\leader$. While there are several important tasks for the government, here we focus on the case where the government is acting to maximise the social welfare of the households. Households are the followers $\followers$, maximising their individual reward, subject to the income tax and asset tax rates from the government.

Each time step $t$, households work $h_t$ for a firm to earn income based on their productivity level $e_t$ and the wage rate $W_t$:

$$
W_t = (1 - \alpha) \biggl(\frac{K_t}{L_t}\biggr)^\alpha
$$
where $\alpha=1/3$ is the capital elasticity, $K_t = \sum a_t$ is the capital, and $L_t = \sum h_t * e_t$ is the aggregate labor. Households consume $c_t$, save $a_t$, and make any required tax payments. The tax payments are set by the government using the HSV tax function \cite{heathcote2017optimal,mi2023taxai}:
\begin{equation}
    T(x, \tau, xi) = x - \frac{1 - \tau}{1-\xi} x^{1-\xi}
\end{equation}
$\tau \geq 0$ controls the average tax rate, and $\xi \geq 0$ the progressivity. Income taxes are $T^\text{i} = T(i_t, \tau^\text{i}, \xi^\text{i})$ and asset taxes $T^\text{a} = T(a_t, \tau^\text{a}, \xi^\text{a})$.

\textbf{Reward}
Households aim to maximise their return, as given by a per timestep reward $r_{\followers,t}$ based on their consumption $c_t$ and working $h_t$ rates. The government aims to maximise the lifetime social welfare for households, receiving per timestep reward $r_{\leader,t}$:

\begin{equation}
\begin{aligned}
    r_{\followers,t} = \log{c_t}  -  \frac{h_t^{1+\zeta}}{1+\zeta}\;, &&
    r_{\leader,t} = \sum_{i \in \followers}  r_{i,t}
    \end{aligned}
\end{equation}

\textbf{Action}
The action for the government $a_{\leader,t}$ is to set the taxation parameters for income and asset taxes. The action for households $a_{\followers, t}$ is to decide their savings ratio $p_t$ and working hours $h_t$:
\begin{equation}
\begin{aligned}
    a_{\leader,t} = \{\tau^{\text{i}}, \xi^{\text{i}}, \tau^{\text{a}}, \xi^{\text{a}} \} = \boldsymbol{\theta}_t, &&
    a_{\followers, t} = \{p_t, h_t\}
\end{aligned}
\end{equation}
where each value is in the unit interval $[0,1]$.

\textbf{Observation}
Households have a view of their current wealth $a_t$, income $i_t$, productivity ability $e_t$, the overall wage rate $W_t$, and the taxation rules set by the government: $$o_{\followers,t} = \{W_t, a_t, i_t, e_t, \boldsymbol{\theta}\}$$
note that here $\boldsymbol{\theta}$ is fully observable, so $\boldsymbol{\theta}=\hat{\boldsymbol{\theta}}_i, \forall i \in \followers$.

Governments have a view of the averages of these values across the households,  $o_{\leader,t} = \{\mathbb{E}[a_t], \mathbb{E}[i_t], \mathbb{E}[e_t], \boldsymbol{\theta}_{t-1}\}$

\subsubsection{Results}

\begin{figure}[!htb]
    \centering
    \includegraphics[width=.8\columnwidth]{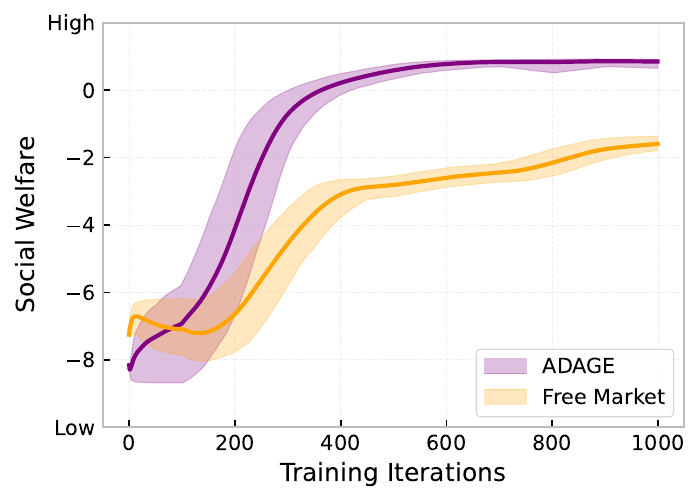}

    \caption{Social welfare throughout training. The solid line shows the mean value, and the filled region shows the entire range (min $\dots$ max).}
    \label{figWelfare}
\end{figure}

We investigate how successfully an adaptive outer layer can design taxation policies to maximise social welfare $R_{\leader}$. For this, we compare to a free market (no tax) baseline. The proposed approach successfully maximises the social welfare in this environment (\cref{figWelfare}), significantly improving upon the free market case, indicating that the adaptive framework can successfully learn policies to improve social welfare in a complex economic system. 

\begin{figure*}[!htb]
     \centering
     \begin{subfigure}[b]{0.3\textwidth}
         \centering
         \includegraphics[width=\textwidth]{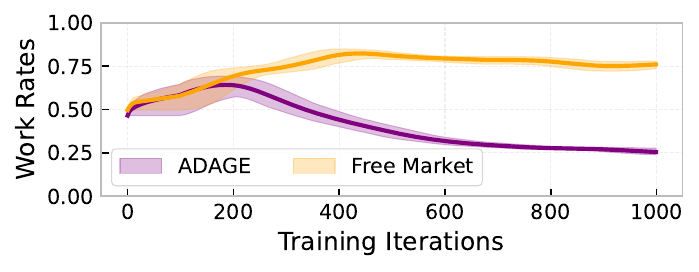}
         \caption{Work rates}\label{figWorkRates}
     \end{subfigure}
     \hfill
     \begin{subfigure}[b]{0.3\textwidth}
         \centering
         \includegraphics[width=\textwidth]{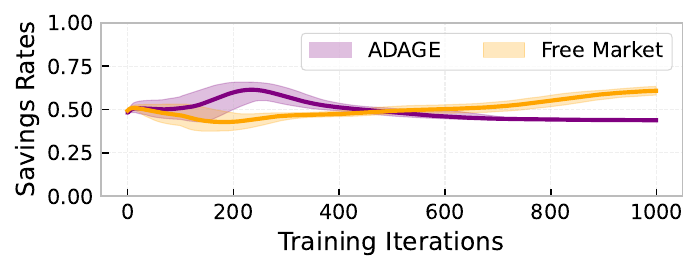}
         \caption{Saving rates}\label{figSavingsrates}
     \end{subfigure}
     \hfill
     \begin{subfigure}[b]{0.3\textwidth}
         \centering
         \includegraphics[width=\textwidth]{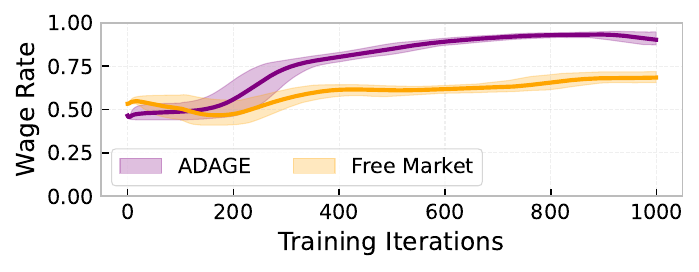}
         \caption{Wage rate}\label{figWagerate}
     \end{subfigure}
        \caption{Policy Design: Resulting mean household work, savings, and wage rates throughout training.}
        \label{figPolicyDesign}
\end{figure*}

\begin{figure}[!htb]
    \centering
\begin{subfigure}[b]{0.4\columnwidth}
         \centering
         \includegraphics[width=\textwidth]{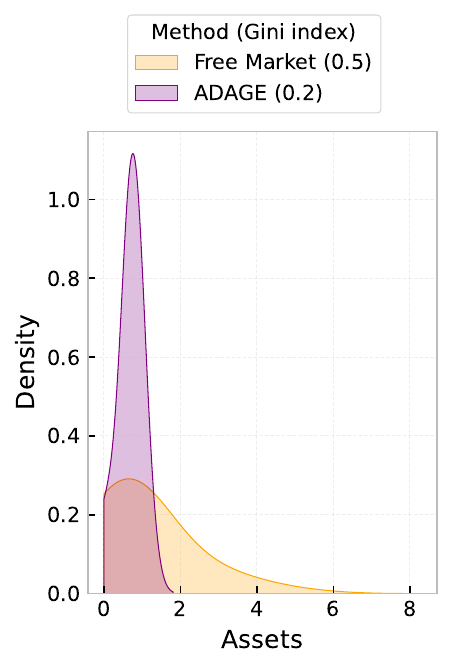}
         \caption{Assets}
         \label{figGiniAssets}
     \end{subfigure}
     \begin{subfigure}[b]{0.4\columnwidth}
         \centering
         \includegraphics[width=\textwidth]{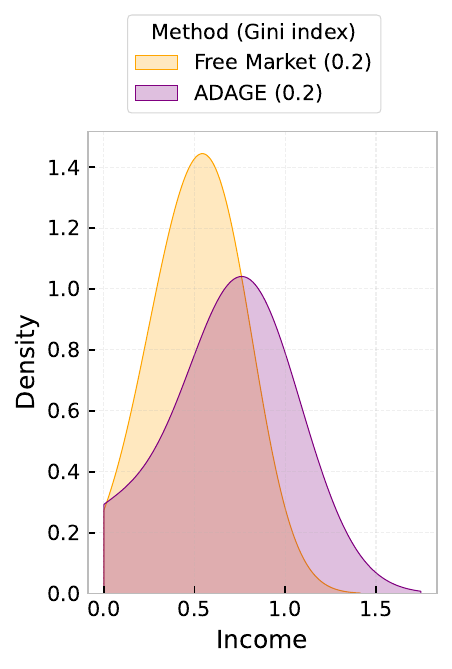}
         \caption{Income}
         \label{figGiniIncome}
     \end{subfigure}
    \caption{Household inequality from multiple rollouts.}
    \label{figGini}
\end{figure}

This improvement in social welfare is achieved through households working less (\cref{figWorkRates}), earning higher wages (\cref{figWagerate}), and consuming more, which comes at the expense of lower savings rates (\cref{figSavingsrates}). As households adjust their behaviour to the taxation rates, higher asset taxes encourage this additional consumption. Additionally, while not directly optimised for, the learnt taxation rules also improve equality in the system (\cref{figGini}), reducing the right-tailed asset distribution, minimising the household asset inequality (\cref{figGiniAssets}) while achieving higher average income rates at no expense to the income equality (\cref{figGiniIncome}), resulting in a more equal population.

\subsection{Calibrator}\label{secCalibration}

Another crucial modelling task is calibration \cite{platt2020comparison}. Often, there are simulation parameters where the exact values are unknown, and must instead be calibrated to match some real-world dynamics to improve the simulator's realism. Examples include financial simulators \cite{dyer2023gradient}, agent compositions \cite{vadori2020calibration}, and models of human behaviour \cite{evans2024learning}. Again, this section demonstrates how calibration naturally fits within the ADAGE framework.

\subsubsection{Environment} 

To demonstrate how calibration can be performed with ADAGE, we use the Cobweb market game \cite{hommes2007learning} to explore market price fluctuations, where with human participants, we see larger fluctuations than is to be expected by perfectly rational participants, so we must calibrate the information processing costs (modulating the bounded rationality) of the agents \cite{evans2024learning}.

In a cobweb market, $i \in \followers$ producers must estimate the price $p_t$ of a good at the next timestep $t$. The price is a result of the estimates $\hat{p}_{i,t}$ from each producer:

\begin{equation}
    p_t = \frac{a - \sum_{j \in n} \mathcal{S}(\hat{p}_{j,t})}{b} + \epsilon_t
\end{equation}
where $\mathcal{S}(\hat{p}_{i,t}) =\tanh (\psi (\hat{p}_{i,t} - n)) + 1$ is the supply curve, and $a=13.8$, $b=1.5$, $\psi=2$ are market specific parameters (from \cite{hommes2007learning}), and $\epsilon$ is a noise term (additional details in \cref{secAppendixCalibrate}).

Producers receive a reward $r_{\followers,t}$ based on the accuracy of their prediction $\hat{p}_{i,t}$ (\cref{eqCobweb}). However, each are boundedly rational, therefore they are subject to an information processing constraint when maximising $R_{\followers}$:
\begin{equation}
\text{ subject to } I(\pi_i, o_{i,t}) < \Bar{I}
\end{equation}
giving learnt follower behavioural policies $i \in \followers$ of the form
\begin{equation}\label{eqRewardMod}
    \pi_i(a_i|o_i, \lambda_i) = \max_{\pi_i}\mathbb{E}_{\pi_i}\left[\sum_{t=0}^\infty\gamma^t \biggl( r_{i,t} (s_t, a_{i, t}, a_{-i, t}) - \lambda_i I(\pi_i, o_{i,t}) \biggr)\right]
\end{equation}
encoding the soft information constraint based on some calibratable parameters $\lambda_i, i \in \followers$, i.e., $\hat{\boldsymbol{\theta}}_i=\lambda_i$. While different information cost functions $I$ can be used, here we use the KL penalty \cite{evans2024bounded,ortega2013thermodynamics}: $$I(\pi, o) = \sum_{a} \pi(a| o) \log \frac{\pi(a| o)}{ \pi_0(a)}$$
where $\pi_0(a)$ is some prior belief, which we assume to be uniform.

\textbf{Reward}
The calibrator $L$ is attempting to find parameters $\boldsymbol{\lambda} = \boldsymbol{\theta}$ to minimise the difference in the distribution of the experimental market prices $\varphi$ and the simulated market prices:
\begin{equation}
    r_{\leader,t} =-|\varphi - p_t|
\end{equation}
using the absolute error per timestep.

Producers receive a non-linear reward based on the accuracy of their prediction compared to the resulting market price $p_t$ \cite{hommes2007learning}:
\begin{equation}\label{eqCobweb}
    r_{\followers,t} = \max(0, 1300-260(p_t - \hat{p}_{i,t})^2)
\end{equation}

\textbf{Action}
The action for the calibrator is a vector action $$a_{\leader,t} = \{\mu, \sigma\} = \boldsymbol{\theta}$$ to decide distributional parameters $\mathcal{N}(\mu, \sigma)$, where individual processing resources $\lambda_i$ are sampled from this distribution $\lambda_i \sim \mathcal{N}(\mu, \sigma)$ (as in \cite{evans2024learning}). While here we demonstrate the case of learning distribution parameters, an alternative approach could also learn the individual $\lambda_i$ directly, as demonstrated in \cref{secAdditionalCalibration}.

\textbf{Observation} The observation for the calibrator at time $t+1$ is the prediction vector of the producers $\boldsymbol{\hat{p}}_{t}$ and the past price $p_t$:
$$
o_{\leader,t+1} = \{ \boldsymbol{\hat{p}}_{t}, p_t \}
$$

The observation for the followers at $t+1$ is the realised market price at $p_t$, the mean historical market price $\mathbb{E}[p_{\leq t}]$, their own mean predicted price $\mathbb{E}[p_{i, \leq t}]$, and their processing penalty $\hat{\boldsymbol{\theta}}_i=\lambda_i$:
$$
o_{\followers, t+1} = \{ \mathbb{E}[p_{\leq t}], \mathbb{E}[p_{i, \leq t}], p_t, \lambda_i \}
$$
note that the producers only have a local view of the parameters $\boldsymbol{\theta}$, only knowing their individual $\lambda_i$.
\subsubsection{Results}

To investigate how successfully the adaptive outer layer calibrates a simulator to real-world data, we compare with uncalibrated baselines, including the analytically derived equilibrium and a base simulator with rational agents ($\lambda_\followers=0$) to show the importance of appropriate calibration. Additionally, while the task is not to compare specific calibration algorithms, we show how other algorithms can be natively integrated into ADAGE in the supplementary material (e.g. Bayesian optimisation in \cref{secAdditionalCalibration}) to show compatibility with the framework.

\begin{table}[]
\caption{Calibration metrics (based on the mean error from multiple rollouts to bootstrapped samples from experimental data of \cite{hommes2007learning}).
}\label{tblCalibration}
\footnotesize
\begin{tabular}{@{}llll@{}}
\toprule
                       & \textbf{Equilibria} & \textbf{Base Simulator} & \textbf{ADAGE} \\ \midrule
\textbf{MAE}           &    0.01                 &               0.013          &       \textbf{0.005}       \\
\textbf{RMSE}           &    0.017                 &            0.026             &             \textbf{0.007}  \\
\bottomrule
\end{tabular}
\end{table}

\begin{figure}[!htb]
    \centering
\includegraphics[width=.8\columnwidth]{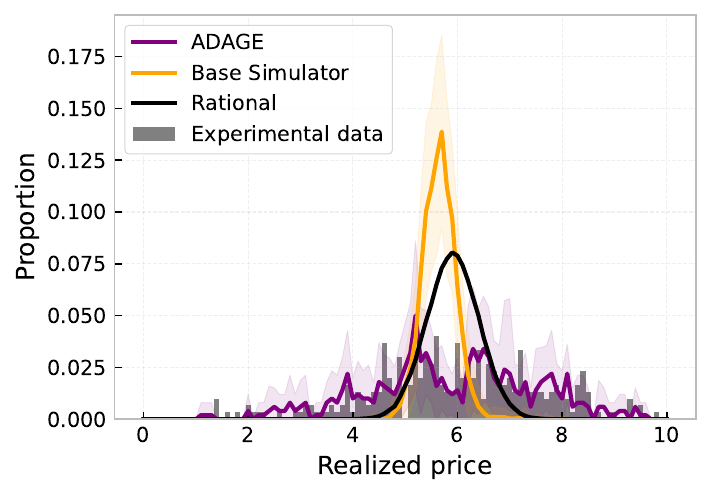}
    \caption{Calibration results (experimental data from \cite{hommes2007learning}) over multiple rollouts.}
    \label{figCalibration}
\end{figure}

The outer layer successfully calibrates the parameters to to the experimental data, 
well capturing the overall distribution (\cref{figCalibration}) and producing the lowest error (\cref{tblCalibration}). These results demonstrate that ADAGE can also perform model calibration and that models of bounded rationality are still fully compatible with the learning-based framework, alleviating concerns about perfectly rational agents. We demonstrated three calibration approaches. One which optimised parameters to a distribution of information processing penalties with RL (\cref{figCalibration}), one which optimised these penalties directly with RL (supplementary material, \cref{secAdditionalCalibration}), and additionally, one which used a Bayesian outer layer to calibrate the parameters (supplementary material, \cref{secAdditionalCalibration}) to show the framework is independent of the exact learning algorithm used.

In contrast to \cref{secPolicyDesign}, which maximised social welfare, here, we calibrated a simulator to real data. Despite the seemingly distinct goals, both tasks can be seen as specific instantiations of ADAGE by changing the reward and action space of the leader agent.

\subsection{Scenario Generator}

Another common modelling task is scenario generation \cite{10.5555/3635637.3662899}, determining under what conditions specific scenarios emerge from a simulation. For example, when do crises form \cite{praiwattana2016survey} or opinions split 
\cite{10.5555/3635637.3662899}. Despite being distinct from the previous tasks, we demonstrate the compatibility with the proposed framework.

\subsubsection{Environment} 
Here, we want to reduce (emergent) market volatility to promote more stable regimes. Therefore, the goal for the outer layer is to generate scenarios for stabilising a market. Specifically, we explore a market entrance game \cite{patrick2023bounded} and the impact of Tobins tax on controlling the volatility in this market \cite{bianconi2009effects,Sanyal2019}. Tobins tax was initially introduced \cite{tobin1978proposal} as an approach to penalising short-term currency conversions to stabilise the foreign exchange market. Here, we explore this idea in a generic market setting. 

The outer layer is the scenario generator and sets the Tobins tax duty $\boldsymbol{\theta}=\tau_t$. The inner layer is composed of $n$ trading agents, making simulatenous decisions on whether or not to enter or stay out of a market, where the utility for the traders depends on the market entrance decisions of the other traders \cite{evans2024bounded}.


\textbf{Reward} The base per timestep utility for the traders is 

\begin{equation}\label{eqRewardMarketEntry}
r^*_{k,t} = 
\begin{cases}
\beta, & \text{if } a_{k,t} = \text{stay out} \\
\beta + \upsilon(C - D_t), & \text{if } a_{k,t}  = \text{enter} \\
\end{cases}
\end{equation}
where $\beta=1$ is the baseline reward (e.g., a risk-free rate if they decided not to enter a market), the demand $D_t = \sum_{k \in \followers} \mathds{1}\{a_{k,t} = \text{enter}\}$ is the total number of traders entering the market, $0 \leq C \leq n$ is the market capacity, and $\upsilon=2$ is the strength of the reward/penalty for risking entering the market. \Cref{eqRewardMarketEntry} rewards entrance with excess market capacity as there is more room to profit due to reduced competition, and penalises entry into overly saturated markets ($D > C$) where profit diminishes.

The reward for the outer layer is the (negative) standard deviation in historical demand, where larger fluctuations (i.e. more volatile markets) are implicitly discouraged:
\begin{equation}
    R_{T, \leader} = - \sigma\big(D_T\big) = - \sqrt{ \frac{\sum_{i=1}^{T} (D_i - \mathbb{E}[D])^2 }{T - 1}}
\end{equation}

\textbf{Action} The action for traders is a binary (discrete) decision on whether or not to enter the market:
$$
A_s = \{\text{enter}, \text{stay out}\}
$$
The action for the outer layer is to set the global tax rate $\boldsymbol{\theta} = \hat{\boldsymbol{\theta}}_i = 0 \leq \tau_t \leq 0.1$. The tax must be paid by the follower agents changing their market positions, giving the final reward $r$ of: 
$$
r_{k,t} = 
\begin{cases}
    r^*_{k,t} - |\tau \times r^*_{k,t} |, & \text{if } a_{k,t} \neq a_{k,t-1} \\
    r^*_{k,t} & \text{otherwise} \\
\end{cases}
$$

\textbf{Observation} Trader $k$ observes the previous demand, the mean historical demand, their previous position, and the tax rate:
$$
o_{k,t} = \{D_{t-1}, \mathbb{E}[D_{<t}], a_{k, t-1}, \hat{\boldsymbol{\theta}}_i\}
$$
The observation for the scenario generator (leader) agent is the running standard deviation of the demand (giving an estimate of the volatility) as well as the current tax rate:
$$
o_{\leader,t} = \{\sigma\big(D_t\big), \tau_t \}
$$

\subsubsection{Results}

\begin{table*}[!htb]
\centering
 \caption{Resulting mean volatility metrics across different entrance capacities $C$ (the lower the \textbf{better}) for multiple rollouts. Timeseries for each capacity are visualised in the supplementary material \cref{figAtendanceExtended}.}\label{tblCResults}
\footnotesize
\begin{tabular}{@{}llllllllll|lllllllll@{}}
\toprule
 & \multicolumn{9}{c}{Mean $\sigma$} & \multicolumn{9}{c}{Mean absolute percentage change} \\ \midrule
\multicolumn{1}{r}{$C$} & 2 & 4 & 6 & 8 & 10 & 12 & 14 & 16 & 18 & 2 & 4 & 6 & 8 & 10 & 12 & 14 & 16 & 18 \\
\textbf{Baseline} & 1.35 & 0.92 & \textbf{1.23} & 1.64 & 3.67 & 2.13 & 1.88 & \textbf{1.39} & 2.22 & 0.66 & \textbf{0.09} & 0.49 & 0.25 & \textbf{0.08} & 0.16 & 0.14 & \textbf{0.09} & 0.15 \\
\textbf{ADAGE} & \textbf{1.20} & \textbf{0.86} & 1.42 & \textbf{1.48} & \textbf{1.49} & \textbf{1.62} & \textbf{1.31} & 1.97 & \textbf{0.75} & \textbf{0.63} & 0.14 & \textbf{0.07} & \textbf{0.19} & 0.13 & \textbf{0.05} & \textbf{0.06} & 0.14 & \textbf{0.04} \\ \bottomrule
\end{tabular}
\end{table*}

\begin{figure}[!htb]
     \centering
     \begin{subfigure}[b]{.8\columnwidth}
         \centering
         \includegraphics[width=\textwidth]{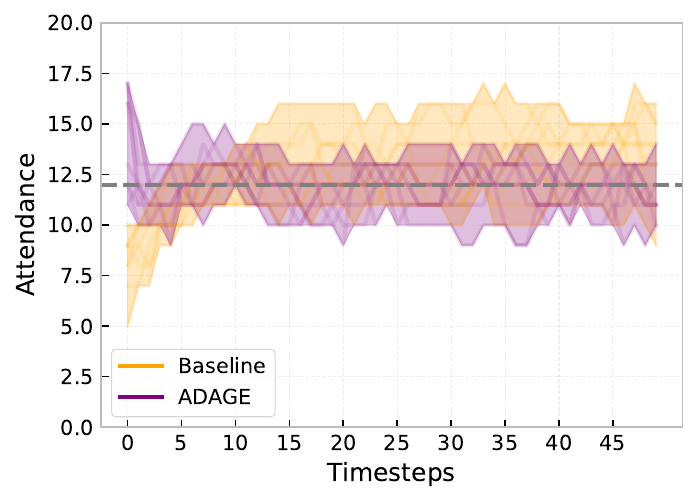}
         \caption{Timeseries}
         \label{figMarketEntryTimeseries}
     \end{subfigure}
     \hfill
     \begin{subfigure}[b]{\columnwidth}
         \centering
         \includegraphics[width=.7\textwidth]{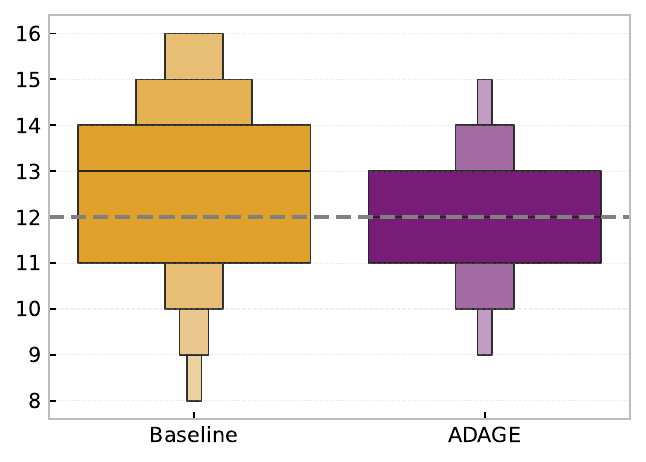}
         \caption{Distributions}
         \label{figMarketEntryDistribution}
     \end{subfigure}
     \hfill
        \caption{Market entrances over multiple rollouts. In each plot, the dashed horizontal line indicates the capacity $C$.}
        \label{figMarketEntrance}
\end{figure}

The proposed approach, which features a tax rate set by the leader policy, is compared to the baseline approach (no Tobins tax), across volatility metrics and capacities in \cref{tblCResults}. The outer layer discovered automatically that scenarios with a Tobins tax successfully restricts the market volatility (reduces market fluctuations) in the vast majority of configurations. This reduction is observed in the standard deviation of the attendance time series (7/9 cases) and the mean absolute percentage change (6/9 cases).

Focusing on the conventional case of $c=0.6$ \cite{arthur2021foundations}, the resulting market timeseries is visualised in \cref{figMarketEntryTimeseries}, showing a far narrower range of attendance fluctuations. This reduction in fluctuations is further reaffirmed by the distribution plots in \cref{figMarketEntryDistribution}, successfully narrowing the distribution of attendance, resulting in a more stable market.  Similar results hold across other capacities (\cref{figAtendanceExtended}).

These results show that ADAGE can be used for scenario generation, generating the desired outcome (in this case, reducing market volatility). Additionally, these results help to confirm the manual findings of \cite{bianconi2009effects, Sanyal2019}, that a Tobins tax can help control market volatility by penalising frequent trading behaviour, which was discovered automatically here through the bi-level optimisation process.

\subsection{Robust Behavioural Learning}\label{secMetaRL}
Finally, another crucial modelling task is learning robust agent behaviour across environment configurations and/or agent preferences, i.e., meta-learning \cite{gerstgrasser2022meta}. This section demonstrates how ADAGE can be used to learn this robust agent behaviour.

\subsubsection{Environment}
We demonstrate robust behavioural learning under a financial environment with a market maker (MM) trading with various Liquidity Takers (LTs), where the MM may have different preferences in terms of maximising market share or profit and loss (PnL), and the goal is to learn generic MM behaviour across these preferences. For example, a MM trying to maximise market share may provide more desirable pricing through minimising the bid-ask spread they are offering, 
whereas a profit-focused MM will have a larger spread to only execute more profitable trades \cite{vadori2022towards}. 

We follow the general formulation of \cite{vadori2022towards} in a simplified environment inspired by \cite{wright2018evaluating}. There is one learning MM agent and $n$ zero-intelligence LT agents \cite{ladley2012zero,farmer2005predictive} interacting in a market. The key focus here is on the behaviour of the MM under various preferences.

Prices in this market are exogenously determined and follow a mean reverting process around a reference (mean) price $p_0$ \cite{wright2018evaluating}: 

\begin{equation}
    p_t = max(0, \kappa p_0 + (1-\kappa) p_{t-1} + \mathcal{N}(0, \sigma^2))
\end{equation}
where $\kappa=0.01$ is the mean reversion and $\sigma$ the shock variance. 

\textbf{Reward} The reward for the MM is a mixture of PnL and market share $m$, given by: 

\begin{equation}
    r_{\text{MM},t} = \omega \frac{PnL_t}{n} + (1 - \omega) \frac{m_t}{n}
\end{equation}
\begin{equation}
\text{PnL}_t =  q^\text{bid}_t \times (p_{t-1} - p^\text{bid}_t) + q^\text{ask}_t \times (p^\text{ask}_t - p_{t-1}),
\end{equation}
\begin{equation}\label{eqMarketMakerReward}
    m_t = q^\text{bid}_t + q^a_t
\end{equation}
where $q^\text{bid}_t$ ($q^\text{ask}_t$ ) is the quantity the MM bought (sold), and $\omega$ is the MMs preference. 

The reward for the outer layer (the meta learner) is to ensure that the MM is exposed to a variety of preferences $\omega \in \boldsymbol{\omega}$ in order to learn robust behaviour, which we encode as the entropy of the sampled preferences:
\begin{equation}\label{eqMetaOuterReward}
    R_{\leader} = -\sum_{\omega \in \boldsymbol{\omega}} \pi_\leader(\omega) \log(\pi_\leader(\omega))
\end{equation}
encouraging all regions of $\boldsymbol{\omega}$ to be explored. This formulation is based on the principle of indifference; however, if \textit{a priori} preference information is known, this could be encoded directly in $R_{\leader}$.

By optimising for \cref{eqMarketMakerReward} with $\omega$ sampled to maximise \cref{eqMetaOuterReward}, we show how ADAGE can perform meta-learning, with the MM learning robust behaviour across $\boldsymbol{\omega}$. It is important to note that $\boldsymbol{\omega}$ should be sufficiently large to allow a range of preferences to be seen during training, ensuring sufficient extrapolation.

\textbf{Action} The action for the MM is to set the half spread of its pricing $a_{\leader,t} = \text{half spread}_t$, determining the bid $p^\text{bid}_t$ and ask $p^\text{ask}_t$ price for which it is willing to buy and sell:
\begin{equation}
    \begin{aligned}
        p^\text{bid}_t = p_t - \text{half spread}_t, &&
        p^\text{ask}_t = p_t + \text{half spread}_t
    \end{aligned}
\end{equation}
with the resulting spread:
\begin{equation}
    \text{spread}_t = p^\text{ask}_t - p^\text{bid}_t 
\end{equation}
The LTs action spaces are: 
$$
A_{\text{LT}} = 
\begin{cases}
    \{\text{buy}, \text{dont act}\}, & \text{if buyer} \\
    \{\text{sell}, \text{dont act}\}, & \text{if seller}
\end{cases}
$$

LT's are randomly assigned to be a buyer or seller (with equal probability) and follow fixed behavioural rules. LTs have a private valuation for the good $v \sim \mathcal{N}(p_0, \varsigma)$, and will buy (sell) a unit quantity of the good iff $v>p^\text{ask}_t$ ($v<p^\text{bid}_t$), and not act otherwise. 

The action space for the outer layer is to sample $\omega = \hat{\boldsymbol{\theta}}_\text{MM}$. As the behavioural policy which optimises \cref{eqMetaOuterReward} is analytically derivable (the maximum entropy probability distribution), we use this distribution directly instead of learning the behaviour. This shows an additional benefit of ADAGE: when specific behaviour can be computed analytically, this can be substituted directly for any of the behavioural policies, and the remaining behaviour will be optimised w.r.t this, again demonstrating independence on the learning algorithm.

\textbf{Observation} The observation space for the MM includes a normalised view of the market price $p_t$ and its preference $\omega$:
\begin{equation}
    O_t = \{p_t / p_0, \omega\}
\end{equation}
including the preference $\omega$ as part of the observations facilitates meta-learning across various preferences.



\subsubsection{Results} We investigate how the MM adapts their behaviour across distinct preferences $\boldsymbol{\omega} = \{0, 0.25, 0.5, 0.75, 1\}$. The proposed approach is trained across $\boldsymbol{\omega}$ (as sampled by the outer layer) and then evaluated on a specific $\omega \in \boldsymbol{\omega}$, and compared to a baseline model which is trained \textit{and} evaluated on a fixed $\omega \in \boldsymbol{\omega}$. If the adaptive model successfully learns to reproduce the fixed $\omega$ behaviour, we can conclude that the MM successfully generalises across $\boldsymbol{\omega}$.

\begin{figure}[!htb]
    \centering
        \begin{subfigure}[b]{.3\columnwidth}
         \centering
         \includegraphics[height=.14\textheight]{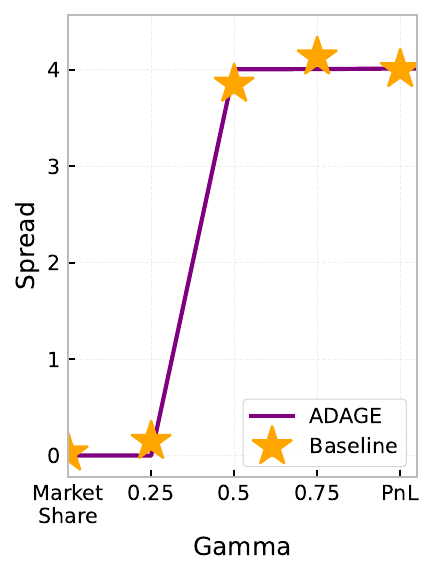}      \caption{Spread}\label{figMetaRLSpread}
     \end{subfigure}
    \begin{subfigure}[b]{.3\columnwidth}
         \centering
         \includegraphics[height=.14\textheight]{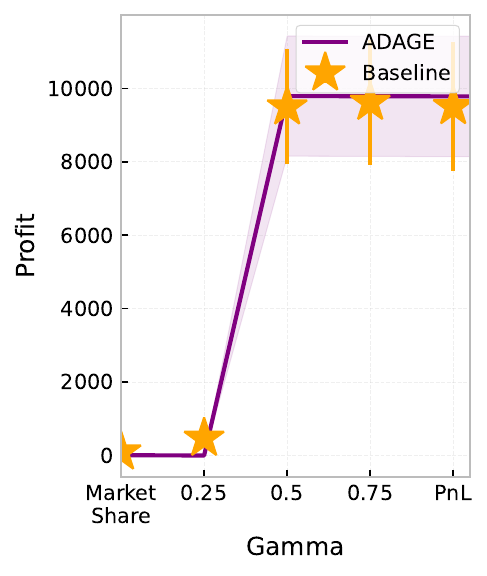}      \caption{Profit}\label{figMetaRLProfit}
     \end{subfigure}
     \begin{subfigure}[b]{.3\columnwidth}
         \centering
         \includegraphics[height=.14\textheight]{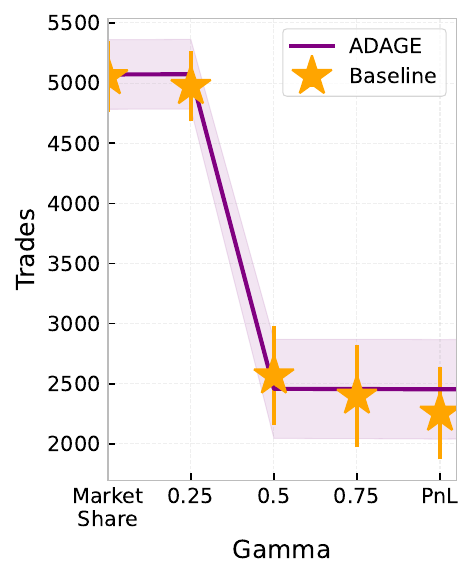}
         \caption{Trades}\label{figMetaRLTrades}
     \end{subfigure}
    \caption{Robust behaviour across $\boldsymbol{\omega}$. The mean for the proposed (baseline) is indicated by the solid line (stars), and the standard deviation by the filled region (vertical star bars) across multiple rollouts.}
    \label{figMetaRL}
\end{figure}

Looking at the resulting spread, profit and market share as a function of $\omega$ (\cref{figMetaRL}), both the baseline and the proposed approach learn appropriate behaviour to balance the reward. When the preference is market share $\omega \to 0$, the MM minimises its spread, maximising the number of trades it conducts by offering more desirable pricing. As the preference shifts to PnL $\omega \to 1$, the MM has a higher spread, only executing the more profitable trades.
Notably, despite having a single behavioural policy, the proposed approach can extrapolate well across $\boldsymbol{\omega}$, reproducing the fixed $\omega$ case which required retraining under each preference (i.e. $5$ distinct behavioural policies, one for each $\omega$). These results show that ADAGE can learn a robust range of behaviours across distinct agent preferences with a single behavioural policy, generalising to the different settings.

\subsection{Key takeaways}
We demonstrated how ADAGE encapsulates common (previously viewed as distinct) agent-based modelling tasks under one unified framework. Specifically, we have shown this for:
\vskip.5\baselineskip
\begin{minipage}{.35\columnwidth}
\begin{itemize}
    \item \textbf{Policy design}
    \item \textbf{Calibration}
\end{itemize}
\end{minipage}
\begin{minipage}{.6\columnwidth}
\begin{itemize}
    \item \textbf{Scenario generation}
    \item \textbf{Robust behavioural learning}
\end{itemize}
\end{minipage}
\vskip.5\baselineskip
\noindent However, these are just example tasks, and many other modelling tasks are possible due to ADAGEs flexibility. Additionally, while we primarily used one particular optimisation algorithm based on RL (PPO, \cref{appendixOpt}), the framework is independent of the exact optimisation algorithm used. For example, in \cref{secMetaRL}, we showed integration of an analytically derived outer layer, and in \cref{secAdditionalCalibration}, we demonstrated a Bayesian outer layer. As the goal was not to analyse the effectiveness of different optimisation algorithms but rather develop a generic framework for adaptive agent-based modelling, we have left out a head-to-head comparison of the possible learning algorithms, and instead shown the integration with different models, tasks, and learning algorithms.

\section{Discussion and Conclusion}
We introduced ADAGE, a novel adaptive agent-based modelling framework in which agents and their environment co-evolve, with agents adjusting their behaviour in response to environmental changes and vice versa. We demonstrated that ADAGE is generic enough to encompass a variety of common agent-based modelling tasks. For example, we used ADAGE to calibrate simulators by treating latent parameters as adjustable characteristics. We explored ADAGEs application in policy design, following frameworks such as the AI Economist \cite{zheng2022ai}, where policies are adjusted to optimize social welfare. Our experiments also highlighted ADAGE's effectiveness in scenario generation, generating desired outcomes, such as stabilizing market dynamics, and, furthermore, in learning robust behavioural rules for ABMs. Given the framework's generality and compatibility with a broad spectrum of ABMs, modelling tasks, and learning algorithms, we hope ADAGE will become a valuable tool in developing adaptive agent-based simulations. Future research could further consider the theoretical aspects of the bi-level optimization process \cite{sihan,thoma2024stochastic}, and explore specialized cases such as differentiable inner layers to refine the framework's adaptability and performance.



\section*{Disclaimer}

This paper was prepared for informational purposes  by the Artificial Intelligence Research group of JPMorgan Chase \& Co. and its affiliates ("JP Morgan'') and is not a product of the Research Department of JP Morgan. JP Morgan makes no representation and warranty whatsoever and disclaims all liability, for the completeness, accuracy or reliability of the information contained herein. This document is not intended as investment research or investment advice, or a recommendation, offer or solicitation for the purchase or sale of any security, financial instrument, financial product or service, or to be used in any way for evaluating the merits of participating in any transaction, and shall not constitute a solicitation under any jurisdiction or to any person, if such solicitation under such jurisdiction or to such person would be unlawful.

© 2025 JPMorgan Chase \& Co. All rights reserved

\bibliographystyle{ACM-Reference-Format} 
\bibliography{bib}

\clearpage

\section*{Supplementary Material}
\appendix

\section{Reinforcement Learning Configuration}\label{appendixOpt}

Behavioural rules in each environment are optimised using Proximal Policy Optimisation with Generalized Advantage Estimation \cite{schulman2017proximal}, as implemented in RLLib \cite{liang2018rllib}. Environments are configured in Phantom \cite{ardon2023phantom}. An overview of the hyperparamaters used are given in \cref{tblParameters}, which are the default values. The training iterations and rollouts are set to allow ample time to converge and enough runs for reducing the resulting stochasticity.

\begin{table}[!htb]
\caption{Algorithm key parameters.}\label{tblParameters}
\begin{tabular}{@{}ll@{}}
\toprule
\textbf{Hyperparamater} & \textbf{Value} \\ \midrule
\textbf{Method} & PPO \\
\textbf{Neural Network Architecture} & [256, 256] \\
\textbf{Activation function} & Tanh \\
\textbf{Learning rate} & 5e-5 \\
\textbf{KL Coeff} & 0.2 \\
\textbf{KL Target} & 0.01 \\
\textbf{Gamma} & 0.99 \\
\textbf{Lambda} & 1 \\
\textbf{Training iterations} & {100, 1000} \\
\textbf{Rollouts} & 10 \\ \bottomrule
\end{tabular}
\end{table}

\section{Additional Discussion}

\subsection{Scenario Generation}
\begin{figure}[!htb]
    \centering
     \begin{subfigure}[b]{\columnwidth}
         \centering
         \includegraphics[width=\textwidth]{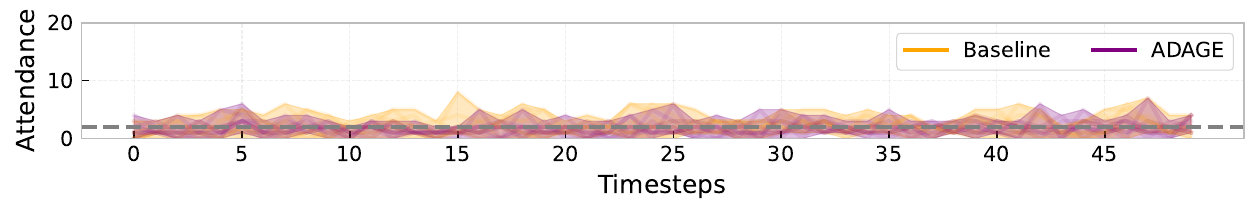}
         \caption{$C=2$}
     \end{subfigure}
     \hfill
     \begin{subfigure}[b]{\columnwidth}
         \centering
         \includegraphics[width=\textwidth]{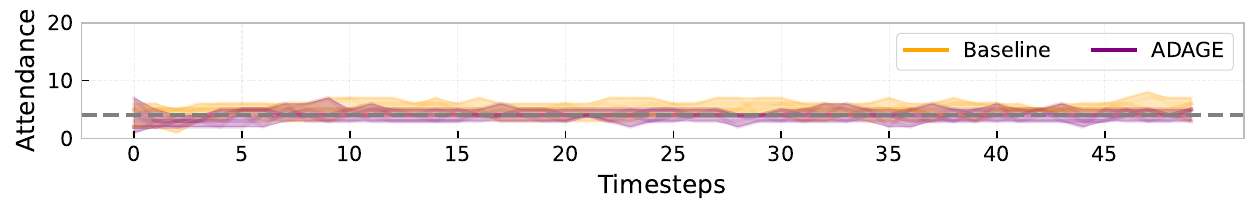}
         \caption{$C=4$}
     \end{subfigure}
     \hfill
     \begin{subfigure}[b]{\columnwidth}
         \centering
        \includegraphics[width=\textwidth]{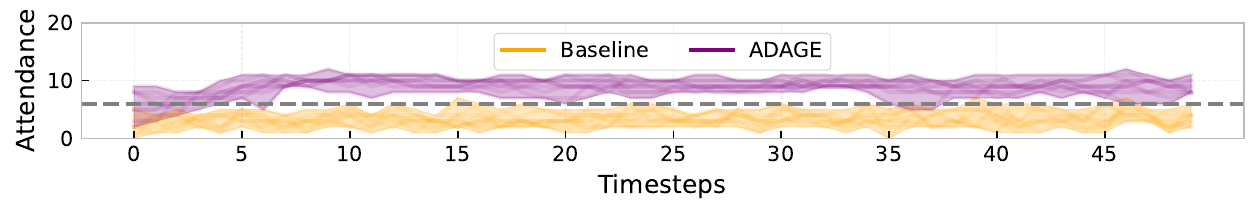}
         \caption{$C=6$}
     \end{subfigure}
          \begin{subfigure}[b]{\columnwidth}
         \centering
         \includegraphics[width=\textwidth]{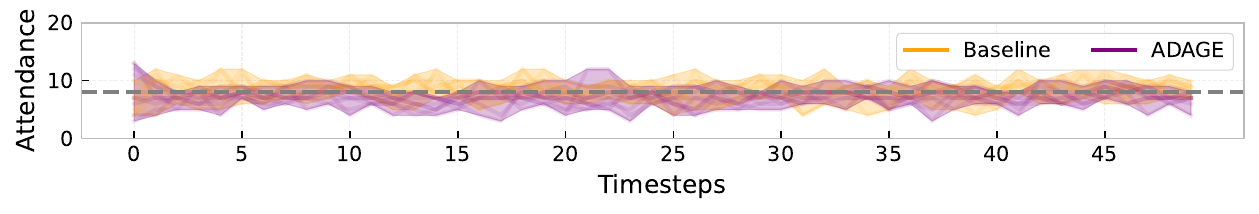}
         \caption{$C=8$}
     \end{subfigure}
     \hfill
     \begin{subfigure}[b]{\columnwidth}
         \centering
         \includegraphics[width=\textwidth]{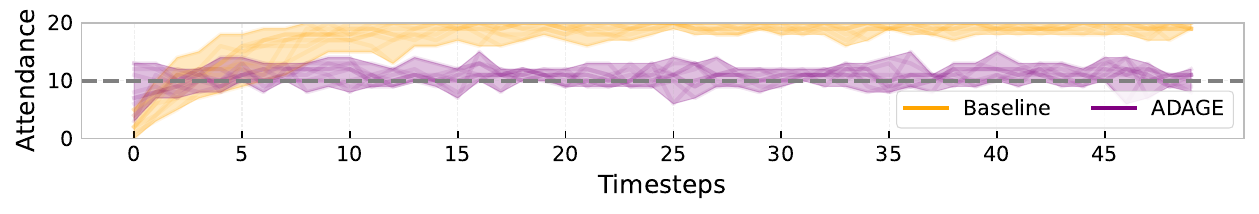}
         \caption{$C=10$}
     \end{subfigure}
     \hfill
          \begin{subfigure}[b]{\columnwidth}
         \centering
         \includegraphics[width=\textwidth]{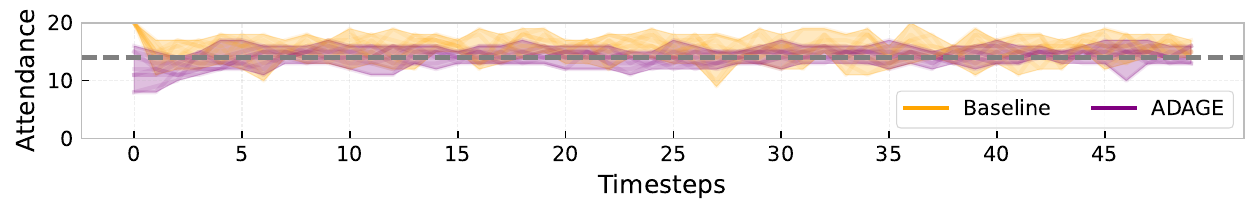}
         \caption{$C=14$}
     \end{subfigure}
     \hfill
     \begin{subfigure}[b]{\columnwidth}
         \centering
         \includegraphics[width=\textwidth]{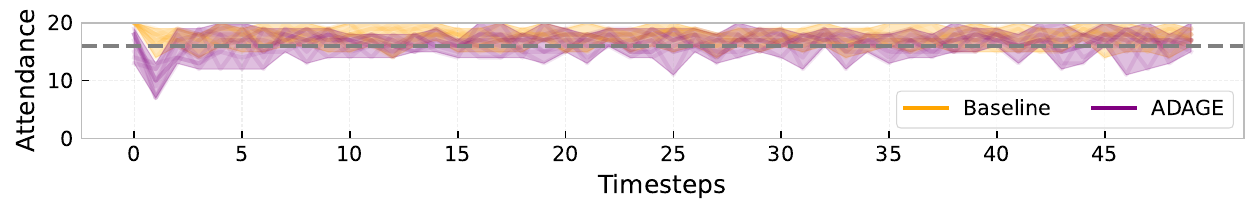}
         \caption{$C=16$}
     \end{subfigure}
     \hfill
     \begin{subfigure}[b]{\columnwidth}
         \centering
         \includegraphics[width=\textwidth]{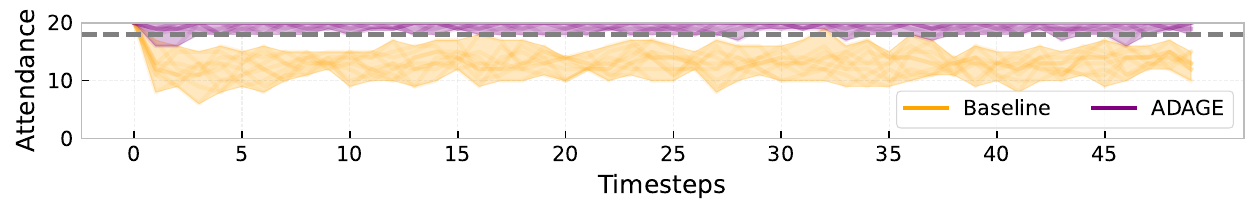}
         \caption{$C=18$}
     \end{subfigure}
        \caption{Attendances across additional entrance capacities}
        \label{figAtendanceExtended}
\end{figure}

To confirm the results are not dependent on the attendance capacity $C$, we evaluate across a broad spectrum of $C$ in \cref{figAtendanceExtended}.
In general, the proposed approach significantly reduces the volatility of attendance fluctuations across various entrance capacities, resulting in more stable markets, supporting our findings.

\subsection{Calibration}

\subsubsection{Extended cobweb market overview}\label{secAppendixCalibrate}
Following \cite{hommes2007learning}, the market price depends on the supply $\mathcal{S}$ and demand $D(p_t) = a - bp_t + \eta_t$ curves. Under rational expectations, producers all predict the price to be the intersection of $\mathcal{S}$ and $D$, $p^*$, e.g., 
$$\bar{p}_{i,t} = p^* + \epsilon_t, \forall i \in K$$meaning the rational predictions will, on average, fall in line with the equilibrium price with fluctuations $\propto \epsilon_t$. 

\subsubsection{Additional results}\label{secAdditionalCalibration}

\begin{figure*}[!htb]
    \centering
    \begin{subfigure}[b]{0.45\textwidth}
    \centering
    \includegraphics[width=\textwidth]{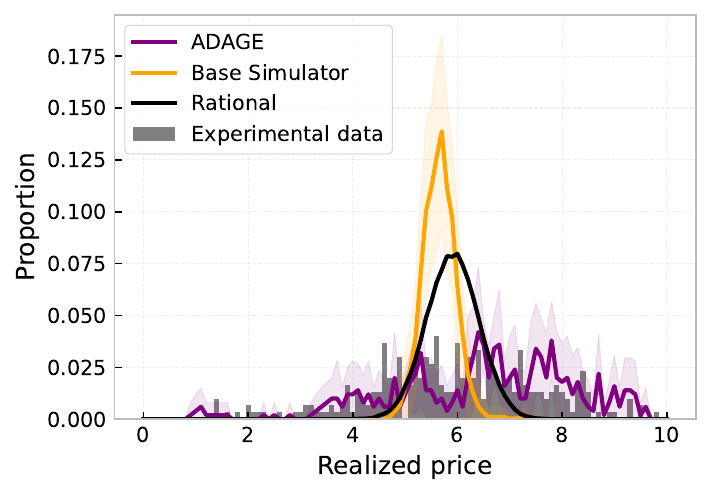}
    \caption{RL Calibration of individual processing resources}
    \label{figCalibrationIndividual}
     \end{subfigure}
     \hfill
     \begin{subfigure}[b]{0.45\textwidth}
    \centering
    \includegraphics[width=\textwidth]{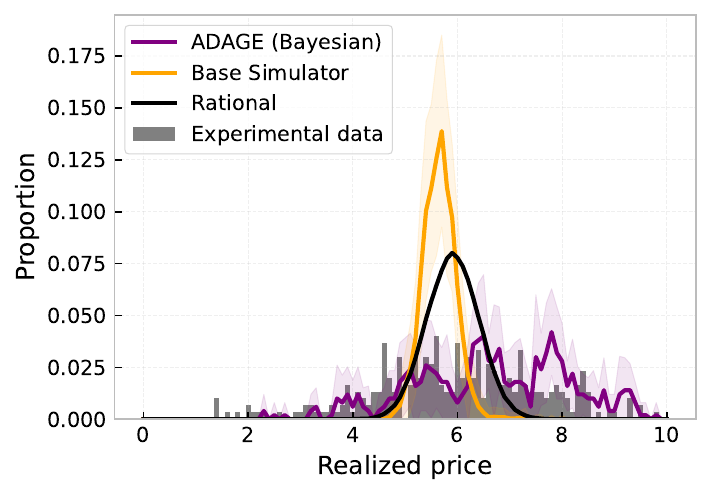}
    \caption{Bayesian calibration}
    \label{figBayesianCalibration}
     \end{subfigure}
\hfill

     \caption{Additional calibration of processing resources plots}
    
\end{figure*}

\textbf{Individual Paramaters} Rather than calibrating distributional parameters, we could instead calibrate the individual parameters for agents, as in \cref{figCalibrationIndividual}. The individual calibration here provides a better fit than the baseline comparisons, but not as good as the distributional parameters. The reason for this reduction when compared to the distributional paramaters is because the underlying experimental results are repeated many times with different participants, so there is naturally a distribution of information processing penalties for each agent, as each agent can represent many different underlying human participants. Nevertheless, this is for illustrative purposes to show it is also possible to calibrate these individual parameters.

\textbf{Bayesian Optimisation} Instead of an RL outer layer, we can substitute this for a Bayesian outer layer. The Bayesian outer layer functions in exactly the same manner as the RL layer, except rather than doing a typical RL update step, each time step performs a Bayesian update step. This is configured with the BayesianOptimisation package \cite{bayesOpt} with default values using Gaussian Processes, and has the same number of iterations and the same parameter bounds as the RL agent. The results for optimising the distribution parameters are shown in \cref{figBayesianCalibration}, demonstrating that while the Bayesian approach integrates well with ADAGE, in this case, the calibration performs worse than the RL settings. However, again, the goal is not to compare the performance between the learning algorithms, but show the independence of ADAGE on the particular learning algorithm. It is likely that with refining the hyperparamaters of the Bayesian we could see improved results, but this is just to demonstrate successful integration.

\end{document}